\documentclass[sigconf,natbib=true]{acmart}

\usepackage{latexsym}
\usepackage{graphicx}
\graphicspath{{./images/}}
\usepackage{booktabs} 
\usepackage{color}  
\usepackage{amsmath}  
\usepackage{subcaption}
\usepackage{caption}
\usepackage{tikz}
\usepackage{colortbl} 
\usepackage{framed}
\usepackage{multirow}
\usepackage{multicol}
\usepackage{hyperref}
\usepackage{url}
\usepackage{balance}
\usepackage{verbatim}
\usepackage{cancel}
\usepackage{svg}
\usepackage{xspace} 
\usepackage[ruled,vlined]{algorithm2e}
\usepackage{bbold} 
\usepackage{centernot}
\usepackage{mathtools}
\usepackage{stmaryrd}
\usepackage{nopageno}

\newcommand{\struct}[1]{\texttt{\small #1}}
\newcommand{\utterance}[1]{\textit{#1}}
\newcommand{\phrase}[1]{\textit{``#1''}}

\newenvironment{Snugshade}[1][236,236,236]{
    \setlength{\itemsep}{0pt}
     \setlength{\parsep}{0pt}
     \setlength{\topsep}{0pt}
     \setlength{\partopsep}{0pt}
     \setlength{\leftmargin}{1.5em}
     \setlength{\labelwidth}{0em}
     \setlength{\labelsep}{0em} 
\setlength{\parskip}{0pt}
    \definecolor{shadecolor}{RGB}{#1}%
    \begin{snugshade}
}{%
    \end{snugshade}%
}

\newcommand{\exaqt}{\textsc{Exaqt}\xspace}
\newcommand{\timequestions}{\textsc{TimeQuestions}\xspace}

\newcommand{\uniqorn}{\textsc{Uniqorn}\xspace}
\newcommand{\graftnet}{\textsc{GRAFT-Net}\xspace}
\newcommand{\pullnet}{\textsc{PullNet}\xspace}
\newcommand{\tagme}{\textsc{TagMe}\xspace}
\newcommand{\elq}{\textsc{ELQ}\xspace}
\newcommand{\aida}{\textsc{AIDA}\xspace}

\newcommand{\myparagraph}[1]{\noindent \textbf{#1}.}

\copyrightyear{2021}
\acmYear{2021}
\setcopyright{rightsretained}
\acmConference[CIKM '21]{Proceedings of the 30th ACM International
Conference on Information and Knowledge Management}{November 1--5,
2021}{Virtual Event, QLD, Australia}
\acmBooktitle{Proceedings of the 30th ACM International Conference on
Information and Knowledge Management (CIKM '21), November 1--5,
2021, Virtual Event, QLD, Australia}

\begin{document}
\pagenumbering{gobble}

\title{Complex Temporal Question Answering on Knowledge Graphs}

\author{Zhen Jia}
\affiliation{%
  \institution{Southwest Jiaotong University, China}
}
\email{zjia@swjtu.edu.cn}

\author{Soumajit Pramanik}
\affiliation{%
  \institution{IIT Bhilai, India}
}
\email{soumajit@iitbhilai.ac.in}

\author{Rishiraj Saha Roy}
\affiliation{%
  \institution{Max Planck Institute for Informatics, Germany}
}
\email{rishiraj@mpi-inf.mpg.de}

\author{Gerhard Weikum}
\affiliation{%
  \institution{Max Planck Institute for Informatics, Germany}
}
\email{weikum@mpi-inf.mpg.de}

\renewcommand{\shortauthors}{Jia et al.}

\newcommand{\squishlist}{
 \begin{list}{$\bullet$}
  { \setlength{\itemsep}{0pt}
     \setlength{\parsep}{1pt}
     \setlength{\topsep}{1pt}
     \setlength{\partopsep}{0pt}
     \setlength{\leftmargin}{1.5em}
     \setlength{\labelwidth}{1em}
     \setlength{\labelsep}{0.5em} } }

\newcommand{\squishend}{
  \end{list}  }

\begin{abstract}

Question answering over knowledge graphs (KG-QA) is a vital topic in IR. Questions with temporal intent are a special class of practical importance, but have not received much attention in research. This work presents \exaqt, the first end-to-end system for answering complex temporal questions that have multiple entities and predicates, and associated temporal conditions. \exaqt answers natural language questions over KGs in two stages, one geared towards high recall, the other towards precision at top ranks. The first step computes question-relevant compact subgraphs within the KG, and judiciously enhances them with pertinent temporal facts, using Group Steiner Trees and fine-tuned BERT models. The second step constructs relational graph convolutional networks (R-GCNs) from the first step's output, and enhances the R-GCNs with time-aware entity embeddings and attention over temporal relations. We evaluate \exaqt on \timequestions, a large dataset of $16k$ temporal questions we compiled from a variety of general purpose KG-QA benchmarks. Results show that \exaqt outperforms three state-of-the-art systems for answering complex questions over KGs, thereby justifying specialized treatment of temporal QA.

\end{abstract}

%
%



\maketitle

\section{Introduction}
\label{sec:introduction}

\begin{figure} [t]
	\centering
	\includegraphics[width=\columnwidth]{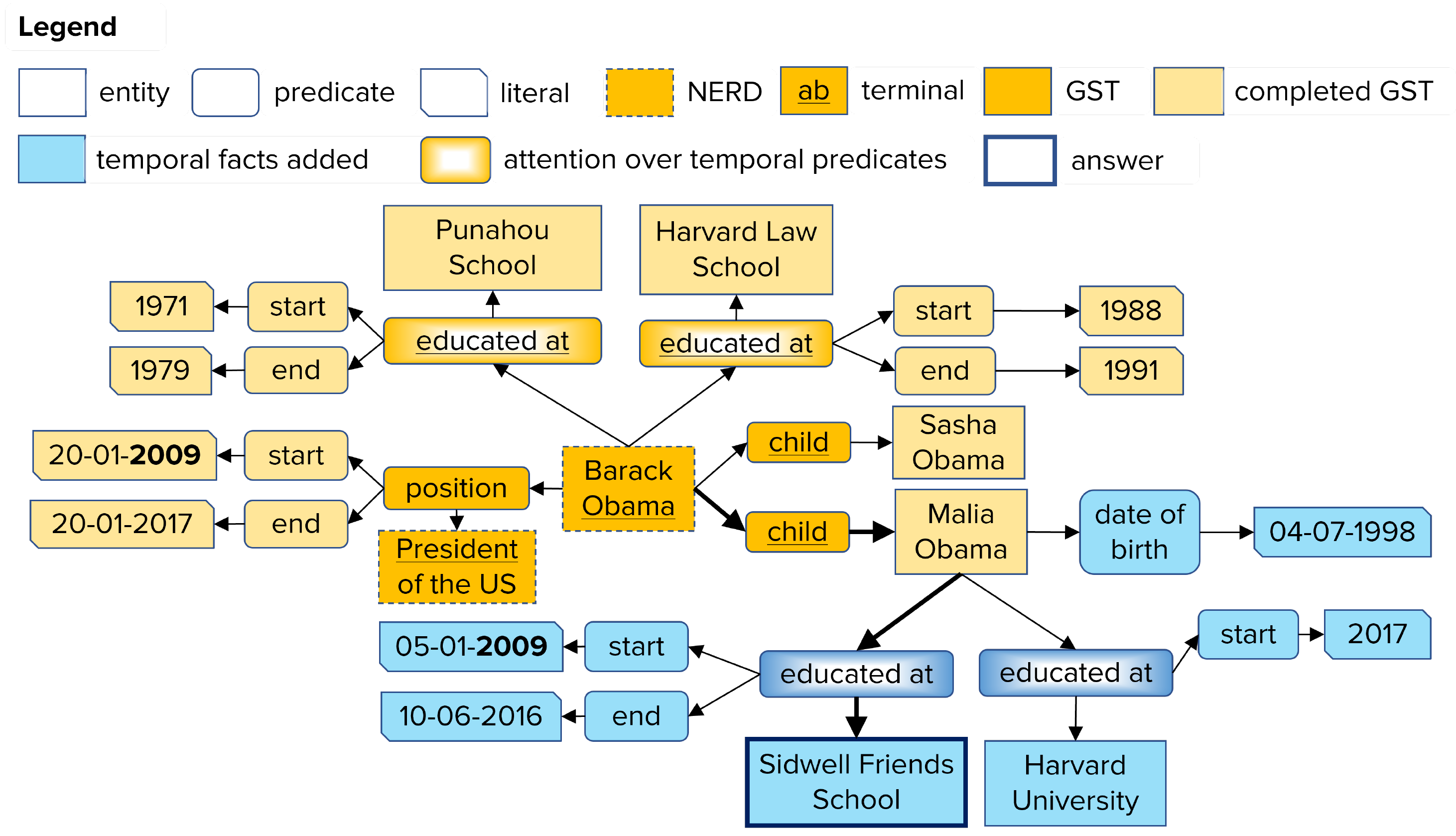}
	\vspace*{-0.7cm}
	\caption{Wikidata excerpt showing the relevant KG zone 
	for the question \utterance{where did obama's children study when he became president?} 
    with answer \struct{Sidwell Friends School}.}
	\label{fig:kg}
	\vspace*{-0.7cm}
\end{figure}

\myparagraph{Motivation} Questions and queries with temporal information needs~\cite{kanhabua2016temporal,alonso2011temporal,alonso2007value,berberich2010language,campos2014survey}
represent a substantial use case in search. For factual questions, knowledge graphs (KGs) like Wikidata~\cite{vrandevcic2014wikidata}, YAGO~\cite{suchanek2007yago},
or DBpedia~\cite{auer2007dbpedia}, 
have 
become
the go-to resource for 
search engines,
tapping into structured facts on entities. 
While question answering over KGs~\cite{bast2015more,bhutani2019learning,saharoy2020question,yahya2013robust,abujabal2018never,vakulenko2019message,berant2013semantic,yih2015semantic,diefenbach2019qanswer}  
has been 
a major topic, 
little attention has been paid to the case of \textit{temporal questions}. Such questions involve explicit or implicit notions of constraining answers by associated timestamps in the KG. 
This spans a spectrum, starting from simpler cases such as
\utterance{when was obama born?},
\utterance{where did obama live in 2001?},
and \utterance{where did obama live during 9/11?}
to more complex temporal questions like:
\begin{Snugshade}
\utterance{where did obama's children study when he became president?}
\end{Snugshade}
Complex
questions
must consider
multi-hop constraints (\struct{Barack Obama $\mapsto$ child $\mapsto$ Malia Obama, Sasha Obama $\mapsto$ educated at $\mapsto$ Sidwell Friends School}), and reason on the overlap of the intersection of 
time points and intervals
(the start of the presidency in 2009 with the study period at the school, 2009 -- 2016). 
A simplified excerpt of the relevant zone in the Wikidata KG necessary for answering the question, is shown in Fig.~\ref{fig:kg}.
This paper addresses these challenges that arise for complex temporal questions.

\myparagraph{Limitations of state-of-the-art}
Early works on temporal QA over unstructured text sources~\cite{harabagiu2005question,ahn2006supporting,schilder2003temporal,bruce1972model,uzzaman2012evaluating,pustejovsky2002multiple,saquete2009enhancing}
involve various forms of question and document parsing, 
but do not
carry over to KGs with structured facts comprised of entities and predicates.
The few works specifically geared for time-aware QA over KGs include
\cite{jia18tequila,DBLP:conf/cikm/CostaGD20,DBLP:journals/fi/WuZLZG20}.
\cite{jia18tequila}
uses a small set of hand-crafted rules for question decomposition and temporal reasoning. This approach
needs human experts for the rules
and does not 
cope with complex questions.
%
%
\cite{DBLP:conf/cikm/CostaGD20}
creates a QA collection for KGs that capture events and their timelines.
A key-value memory network in~\cite{DBLP:journals/fi/WuZLZG20} includes time information from KGs for answering simple questions.


\myparagraph{Approach} 
We present
\exaqt: \underline{EX}plainable \underline{A}nswering of complex \underline{Q}uestions with \underline{T}emporal intent,
a system that does not rely on manual rules for question understanding and 
reasoning. \exaqt answers complex temporal questions
in two steps: 
\squishlist
\item[(i)] Identifying a compact, tractable \textit{answer graph} that contains all 
cues required for 
answering the question, based on
dense-subgraph algorithms and fine-tuned BERT models;
and
\item[(ii)] A \textit{relational graph convolutional network (R-GCN)}~\cite{sun2018open}
to infer the answer in the graph, augmented with signals about time.
\squishend
%
The two stages work as follows
(partly illustrated in Fig.~\ref{fig:kg}).

\myparagraph{Stage 1: Answer graph construction}
\exaqt fetches all KG facts of entities mentioned in the question (\struct{Barack Obama, President of the United States}: dashed outline boxes), as detected by off-the-shelf NERD systems~\cite{ferragina2010tagme,hoffart2011robust,li2020efficient}. 
The resulting noisy set of facts is distilled into
a tractable
set
by means of a fine-tuned BERT model (admitting information about the children Malia and Sasha, but not Michelle Obama).
To construct a KG subgraph of all question-relevant KG {items} and their interconnections
from this set,
Group Steiner Trees (GST)~\cite{shi2020keyword,lu2019answering,chanial2018connectionlens} are \textit{computed} 
(dark orange nodes, terminals or keyword matches underlined: \phrase{obama}, \phrase{president}, \phrase{child}, \phrase{educated at})
and \textit{completed} (light orange nodes). 
The last and decisive step at this point
augments this candidate answer graph with pertinent \textit{temporal facts}, to bring in cues (potentially multiple hops away from the question entities) 
about relevant dates, events and time-related predicates.
To this end, we
use an analogous BERT model for identifying question-relevant temporal facts
(blue nodes: educational affiliations of
Malia and Sasha and their dates).
The resulting \textit{answer graph} is 
the input of
the second stage.

\myparagraph{Stage 2: Answer prediction by R-GCN}
Inspired by the popular \graftnet model~\cite{sun2018open} and related work~\cite{schlichtkrull2018modeling,sun2019pullnet}, we construct an R-GCN that learns entity embeddings over the 
answer graph and casts answer prediction into a node classification task.
However, R-GCNs as used in prior works are ignorant of temporal constraints~\cite{allen1983maintaining}.
To overcome this obstacle, we augment the R-GCN with time-aware entity embeddings, attention over temporal relations, 
and encodings of timestamps~\cite{zhang2020temporal}, temporal signals~\cite{setzer2002temporal}, and temporal question categories~\cite{jia18tequila}. In our running example, temporal attention helps \exaqt focus on \struct{educated at} as 
a question-relevant relation
(partly shaded nodes). The time-enhanced representation of \struct{Barack Obama} flows through the 
R-GCN (thick edges) and boosts the likelihood of \struct{Sidwell Friends School} as the answer (node with thick borders), which contains 2009 (in {bold}) 
among its
temporal facts.
By producing such concise 
KG snippets for each question (as colored in Fig.~\ref{fig:kg}), \exaqt yields explainable evidence 
for its answers.

\myparagraph{Contributions}
This work makes the following contributions:
\squishlist
    \item We propose \exaqt, the first end-to-end system for answering complex temporal questions over large-scale knowledge graphs;
    \item \exaqt 
    applies fine-tuned BERT models and convolutional graph networks to solve the specific challenges of identifying relevant KG facts for complex temporal questions;
    \item We compile and release \timequestions, a benchmark of about $16k$ temporal questions (examples in Table~\ref{tab:qegs});
    \item 
    Experiments over the full Wikidata KG show the superiority of \exaqt over three state-of-the-art complex KG-QA baselines. All resources from this project are available at \url{https://exaqt.mpi-inf.mpg.de/} and \url{https://github.com/zhenjia2017/EXAQT}.
\squishend

\section{Concepts and notation}
\label{sec:concepts}


\begin{table} [t]
	\centering
	\resizebox*{\columnwidth}{!}{
		\begin{tabular}{l l}
			\toprule
			\textbf{Category}   &   \textbf{Question}	                                                            \\  \toprule
			                    &   \utterance{who won oscar for best actress 1986?}                                \\
			Explicit            &   \utterance{which movie did jaco van dormael direct in 2009?}                    \\
			                    &   \utterance{what currency is used in germany 2012?}                              \\  \midrule
			        	        &   \utterance{who was king of france during the ninth crusade?}                    \\
			Implicit            &   \utterance{what did thomas jefferson do before he was president?}               \\
			                    &   \utterance{what club did cristiano ronaldo play for after manchester united?}   \\  \midrule
			                    &   \utterance{what was the first film julie andrews starred in?}                   \\
			Ordinal				&   \utterance{what was the second position held by pierre de coubertin?}           \\
			                    &   \utterance{who is elizabeth taylor's last husband?}                             \\  \midrule
				                &   \utterance{what year did lakers win their first championship?}                  \\  
			Temp. Ans.          &   \utterance{when was james cagney's spouse born?}                                \\  
				                &   \utterance{when was the last time the orioles won the world series?}            \\  \bottomrule
	\end{tabular}}
	\caption{Sample temporal questions from \timequestions.}
	\label{tab:qegs}
	\vspace*{-0.7cm}
\end{table}

We now define the 
salient concepts 
that underlie \exaqt.

\myparagraph{Knowledge graph} A knowledge graph (aka knowledge base)
is a collection of facts $F$ organized as a set of \struct{<subject, predicate, object>} triples. It can be stored as an RDF database of such triples, or equivalently as a graph with
nodes
and
edges.
Examples are Wikidata~\cite{vrandevcic2014wikidata}, YAGO~\cite{suchanek2007yago}, DBpedia~\cite{auer2007dbpedia}, Freebase~\cite{bollacker2008freebase} and industrial KGs. When stored as a graph, edges are directed: \struct{subject $\mapsto$ predicate $\mapsto$ object}. 
Subjects and objects are always nodes, while predicates (aka relations) often become edge labels.

\myparagraph{Fact} A fact $f \in F$ can either be binary, containing a subject and an object connected by a predicate, or $n$-ary, combining multiple items via main predicates and qualifier predicates.
An example of a binary fact is \struct{<Barack Obama, child, Malia Obama>}, where subjects are entities (\struct{Barack Obama}), and objects may be entities (\struct{Malia Obama}), literals 
(constants such as dates  
in \struct{<Malia Obama, date of birth, 04-07-1998}>),
or types aka classes
(\struct{private school} in 
\struct{<Sidwell Friends School, type, private school>}). 
We use the terms \textbf{predicate and relation interchangeably} in this text.

An $n$-ary fact 
combines several triples that belong together, 
such as \struct{<Barack Obama, position held, President of the US; start date, 20-01-2009; end date, 20-01-2017>} 
(see Fig.~\ref{fig:kg}).
\struct{position held} is the main predicate, \struct{President of the US} is the main object, while the remaining data are <qualifier predicate, qualifier object> pairs. $n$-ary facts are of vital importance in temporal QA, with a large fraction of temporal information in modern KGs being stored as qualifiers. 
One way of representing qualifiers in a KG 
is shown in Fig.~\ref{fig:kg}, via paths from the main predicate to the qualifier predicate and on to the qualifier object.
%


\myparagraph{Temporal fact} We define a temporal fact $tf \in F$ as one where the main object or any of the qualifier objects is a timestamp. 
Examples are 
\struct{<Vietnam War, end date, \textbf{30-04-1975}>} (binary),
or,
\struct{<Steven Spielberg, award received, Academy Award for Best Director; for work, Schindler's List; point in time, \textbf{1993}>} ($n$-ary).

\myparagraph{Temporal predicate} We define a temporal predicate as one that can have a timestamp as its direct object or one of its qualifier objects.
Examples are
\struct{date of birth} 
and \struct{position held}. 

\myparagraph{Temporal question} A temporal question is one that contains a temporal expression or a temporal signal, or whose answer is
of temporal nature~\cite{jia2018tempquestions}. Examples of temporal expressions are \phrase{in the year 1998}, \phrase{Obama's presidency},
\phrase{New Year's Eve}, etc. which indicate explicit or implicit temporal scopes~\cite{kuzey2016time}. Temporal signals~\cite{setzer2002temporal} are markers of temporal relations (\struct{BEFORE, AFTER, OVERLAP, ...})~\cite{allen1983maintaining} and are expressed with words like \phrase{prior to, after, during, ...} that indicate the need for temporal reasoning. 
In our models, a
question $q$ is represented as a set of keywords <$q_1, q_2, \ldots q_{|q|}$>.

\myparagraph{Temporal question categories} Temporal questions fall into four basic categories~\cite{jia2018tempquestions}: (i) containing explicit temporal expressions (\phrase{in 2009}), (ii) containing implicit temporal expressions (\phrase{when Obama became president}), (iii) containing temporal ordinals (\phrase{\underline{first} president}), and (iv) having temporal answers (\phrase{When did ...}). 
 Table~\ref{tab:qegs} gives
 several
 examples
 of temporal questions. 
 A question may belong to multiple categories. For example, \utterance{what was the \underline{first} film julie andrews starred in \underline{after her divorce with tony walton}?} contains both an implicit temporal expression and a temporal ordinal.

\myparagraph{Answer} An answer to a temporal question is a (possibly singleton) set of entities or literals,
e. g., \struct{\{Chicago University Lab School, Sidwell Friends School\}}
for \utterance{Where did Malia Obama study before Harvard?}, or \struct{\{08-2017\}} for \utterance{When did Malia start at Harvard?}

\myparagraph{Answer graph} An answer graph is a subset of the KG that contains all the necessary facts for 
correctly answering the question.


\begin{figure} [t]
	\includegraphics[width=\columnwidth]{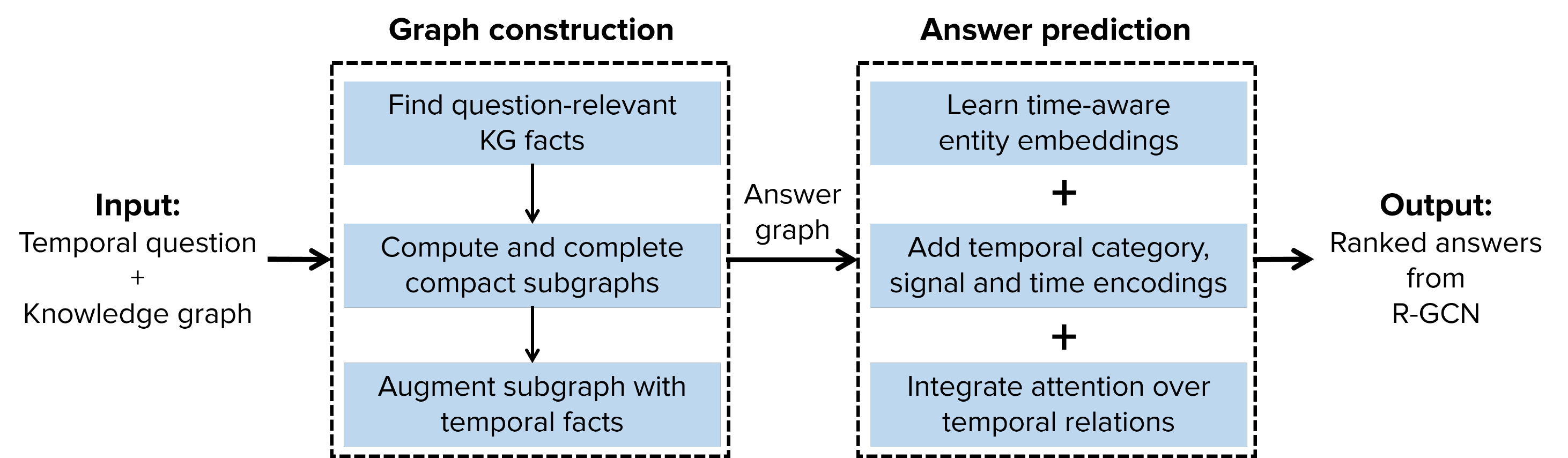}
	\caption{An overview of the two-stage \exaqt pipeline.}
	\label{fig:overview}
	\vspace*{-0.3cm}
\end{figure}

\section{Constructing answer graphs}
\label{sec:method-1}

Fig.~\ref{fig:overview} is an overview of \exaqt, with two main stages: 
(i) answer graph construction (Sec.~\ref{sec:method-1}), and (ii) answer prediction (Sec.~\ref{sec:method-2}). 

\subsection{Finding question-relevant KG facts}
\label{subsec:phase1-1}

\myparagraph{NERD for question entities}
Like most QA pipelines~\cite{qiu2020stepwise,bhutani2019learning}, we start off by running named entity recognition and disambiguation (NERD)~\cite{li2020efficient,van2020rel,hoffart2011robust} on the input question (\utterance{where did obama’s children study when he became president?}). NERD systems identify spans of words in the question as mentions of entities (\phrase{obama}, \phrase{president}), and link these spans to KG items or Wikipedia articles (which can easily be mapped to popular KGs). The facts of these linked entities (\struct{Barack Obama, President of the United States}) provide us with a zone in the KG to start looking for the answer. 
NERD is a critical cog in the QA wheel: entity linking errors leave the main QA pipeline helpless with respect to answer detection.
To mitigate this effect, we  use two different systems, \tagme and \elq \cite{ferragina2010tagme,li2020efficient}, to boost answer recall.
Complex questions often contain multiple entity mentions, and accounting for two NERD systems, we could easily have $2-4$ different entities per question. The total number of associated facts can thus be several hundreds or more.
To reduce this large and noisy set of facts to a few question-relevant ones,
we fine-tune BERT~\cite{devlin2019bert} as follows.

\myparagraph{Training a classifier for question-relevant facts}
For each question in our training set, we run NERD and retrieve all KG facts of the detected entities. We then use a \textit{distant supervision} mechanism: out of these facts, the ones that contain the gold answer(s) are labeled as \textit{positive instances}.
While several complex questions may not have their answer in the facts of the question entities (multi-hop cases), the ones that do, comprise a reasonable amount of training data for our classifier for question-relevance. Note that facts with qualifiers are also retrieved for the question entities (complete facts where the question entity appears as a subject, object, or qualifier object): this increases our coverage for obtaining positive examples.

For each positive instance, we randomly sample five \textit{negative instances} from the facts that do not contain the answer. Sampling question-specific negative instances helps learn a more discriminative classifier, as all negative instances are guaranteed to contain at least one entity from the question (say, \struct{<Barack Obama, spouse, Michelle Obama>}). Using \textit{all facts} that do not contain an answer would result in severe class imbalance, as this is much higher than the number of positive instances. 

We then
pool together the <question, fact> paired positive and negative instances for all training questions.
The fact in this pair is now \textit{verbalized} as a natural language sentence by concatenating its constituents; qualifier statements are joined using \phrase{and}~\cite{oguz2021unified}. 
For example, the full fact for Obama's marriage (a negative instance) is: \struct{<Barack Obama, spouse, Michelle Obama; start date, 03-10-1992; place of marriage, Trinity United Church of Christ>}. This has two qualifiers, and would be verbalized as \phrase{Barack Obama spouse Michelle Obama and start date 03-10-1992 and place of marriage Trinity United Church of Christ.}.
The questions paired with the verbalized facts, along with the 
binary ground-truth
labels, are fed as training input to a \textit{sequence pair classification model} for BERT.

\myparagraph{Applying the classifier}
Following~\cite{devlin2019bert}, the question and the fact are concatenated with the special separator token \struct{[SEP]} in between, and the special classification token \struct{[CLS]} is added in front of this sequence. The final hidden vector corresponding to \struct{[CLS]}, denoted by $\boldsymbol{C} \in \mathbb{R}^H$ ($H$ is the size of the hidden state), is considered to be the accumulated representation. Weights $\boldsymbol{W}$ of a classification layer are the only parameters introduced during fine-tuning, where $\boldsymbol{W} \in \mathbb{R}^{K \times H}$, where $K$ is the number of class labels ($K = 2$ here, fact is question-relevant or not). $\log(\text{softmax}(\boldsymbol{CW^T}))$ is used as the classification loss function. Once the classifier is trained, given a new <question, fact> pair, it outputs the probability (and the label) of the fact being relevant for the question. We make this prediction for all candidate facts pertinent to a question, and sort them in descending order of this question relevance likelihood. We pick the top scoring facts $\{f_{qrel}\}$ from here as our question-relevant set.

\vspace*{-0.3cm}
\subsection{Computing compact subgraphs}
\label{subsec:phase1-2}

The 
set of facts $\{f_{qrel}\}$ contains question-relevant facts but is not indicative as to which are a set of \textit{coherent KG items} that matter for this question, and how they are connected.
To this end, we induce a graph as shown in Fig.~\ref{fig:kg}, from the above set of facts where each KG item (entity, predicate, type, literal) becomes a node of its own. Edges run between components of the same fact in the direction mandated in the KG: \struct{subject $\mapsto$ predicate $\mapsto$ object} for the main fact, and \struct{subject $\mapsto$ predicate $\mapsto$ qualifier predicate $\mapsto$ qualifier object} for (optional) qualifiers.

\myparagraph{Injecting connectivity} BERT selects $\{f_{qrel}\}$ from the facts of a number of entities as detected by our NERD systems. These entities may not be connected to each other via shared KG facts. However, a connected graph is needed so that our subsequent GST and R-GCN algorithms can produce the desired effects. To inject connectivity in the graph induced from BERT facts, we compute the shortest KG path between every pair of question entities, and add these paths to our graph. In case of multiple paths of same length between two entities, they are scored for question-relevance as follows. A KG path is set of facts: a path of length one is made up of one fact (\struct{Barack Obama} $\mapsto$ \struct{position held} $\mapsto$ \struct{President of the United States}), a path of length two is made up of two facts
(\struct{Barack Obama} $\mapsto$ \struct{country} $\mapsto$ \struct{United States of America} $\mapsto$ \struct{office held by head of state} $\mapsto$ \struct{President of the United States}),
and so on. Each candidate path is verbalized as a set of facts (a period separating two facts) and encoded with BERT~\cite{kaiser2021reinforcement}, and so is the question. These BERT encodings are stored in corresponding \struct{[CLS]} tokens. We compute the cosine similarity of \struct{[CLS]}(question) with \struct{[CLS]}(path), and add the path with the highest cosine similarity to our answer graph.

\myparagraph{GST model}
Computing \textit{Group Steiner Trees (GST)}~\cite{shi2020keyword,sun2021finding,pramanik2021uniqorn,lu2019answering} has been shown to be an effective mechanism in identifying query-specific backbone structures in larger graphs, for instance, in keyword search over database graphs~\cite{aditya2002banks,ding2007finding}. Given a subset of nodes in the graph, called \textit{terminals}, the Steiner Tree (ST) 
is the lowest-cost
tree that connects all terminals. This reduces to the minimum spanning tree problem when all nodes of the graph are terminals, and to the shortest path problem when there are only two terminals. The GST models a more complex situation where the terminals are arranged into groups or sets, and it suffices to find a Steiner Tree that connects \textit{at least} one node from each group. This scenario fits our requirement perfectly, where each question keyword can match multiple nodes in the graph, and naturally induces a \textit{terminal group}. Finding a 
tree
that runs through each and every matched node is unrealistic, hence the group model. 

\myparagraph{Edge costs}
An integral part of the GST problem is how to 
define 
\textit{edge costs}. Since edges emanate from 
KG facts, we leverage question-relevance scores assigned by the classifier of Sec.~\ref{subsec:phase1-1}: 
$BERT(f_{qrel}) \in [0, 1]$, converted to  
edge costs $1 - BERT(f_{qrel}) \in [0, 1]$. 

\myparagraph{GST algorithm}
There 
are good
approximation algorithms for GSTs~\cite{li2016efficient,sun2021finding}, but QA needs high precision. Therefore, we adopted the fixed-parameter-tractable exact algorithm by Ding et al.~\cite{ding2007finding}. It iteratively grows and merges smaller trees over the bigger graph to arrive at the minimal trees. Only taking the best tree can be risky in light of spurious connections potentially irrelevant to the question. Thus, we used a top-$k$ variant that is naturally supported by the dynamic programming algorithm of~\cite{ding2007finding}.

\myparagraph{GST completion}
As shown in Fig.~\ref{fig:kg}, the GST yields a skeleton connecting the most relevant question nodes. To transform this into a coherent context for the question, we need to \textit{complete} it with facts from where this skeleton was built. Nodes introduced due to this step are shown in light orange in the figure: dates about the presidency, Obama's children, and the (noisy) fact about Obama's education. In case the graph has multiple connected components
(still possible as our previous connectivity insertions worked only pairwise over entities),
top-$k$ GSTs are computed for each component and the union graph is used for this fact completion step.

\myparagraph{Example} We show a simplified example in Fig.~\ref{fig:kg}, where the node \struct{Barack Obama} matches the question keyword \phrase{Obama}, \struct{child} matches \phrase{children}, \struct{educated at} matches \phrase{study}, and \struct{President of the United States} matches \phrase{president}. The \struct{educated at} nodes connected to Malia and Sasha do not feature here as they are 
not contained in the facts of 
Barack Obama, and do not yet feature in our answer graph. We consider exact matches, although not just in node labels but also in the set of \textit{aliases present in the KG} that list common synonyms of entities, predicates and types. This helps us consider relaxed matches without relying on models like word2vec~\cite{mikolov2013distributed} or GloVe~\cite{pennington2014glove}, that need inconvenient thresholding on similarity values as a noisy proxy for synonyms. The GST is shown using dark orange nodes with the associated question keyword matches underlined (denoting the terminal nodes). In experiments, we only consider as terminals NERD matches for entities, and keyword matches with aliases for other KG items. The GST naturally includes the internal nodes and edges necessary to connect the terminals. Note that the graph is considered \textit{undirected} (equivalently, bidirectional) for the purpose of GST computation.

\vspace*{-0.2cm}
\subsection{Augmenting subgraphs with temporal facts}
\label{subsec:phase1-3}

The final step towards the desired answer graph is to enhance it with temporal facts.
Here, we add \textit{question-relevant temporal facts} of entities in the \textit{completed GST}. This
pulls in temporal information necessary for answering questions that need evidence more than one hop away from the question entities (blue nodes in Fig.~\ref{fig:kg}): \struct{<Malia Obama, educated at, Sidwell Friends School; start date, 05-01-2009>} (+ noise like Malia's date of birth). The rationale behind this step is to 
capture
facts necessary for faithfully answering the question, where faithful refers to arriving at the answer not by chance but after satisfying all necessary constraints in the question. For example, the 
question \utterance{which oscar did leonardo dicaprio win in 2016?} can be answered without temporal reasoning, as he only won one Oscar. We wish to avoid such cases in faithful answering.

To this end,
we first retrieve from the KG all temporal facts of each entity in the completed GST.
We then use 
an analogously 
fine-tuned BERT model for \textit{question-relevance of temporal facts}.
The model predicts, for each temporal fact, its likelihood of containing the answer.
It is trained using temporal facts of question entities that contain the answer as positive examples, while negative examples are chosen at random from these temporal facts. To trap multi-hop temporal questions in our net, we explore 2-hop facts of question entities for ground truth answers. A larger neighborhood was not used during the first fine-tuning as the total number of facts in two hops of question entities is rather large, but the count of 2-hop temporal facts is a much more tractable number. Moreover, this is in line with our focus on complex \textit{temporal} questions.
Let the likelihood score for a temporal fact $tf$ of an entity in the completed GST be
$BERT(tf_{qrel})$.
As before, we take the top scoring
$\{tf_{qrel}\}$,
add them to the answer graph, that is
then passed on to Stage 2. 

\section{Predicting answers with R-GCN}
\label{sec:method-2}

\begin{figure*} [t]
	\centering
	\includegraphics[width=0.95\textwidth]{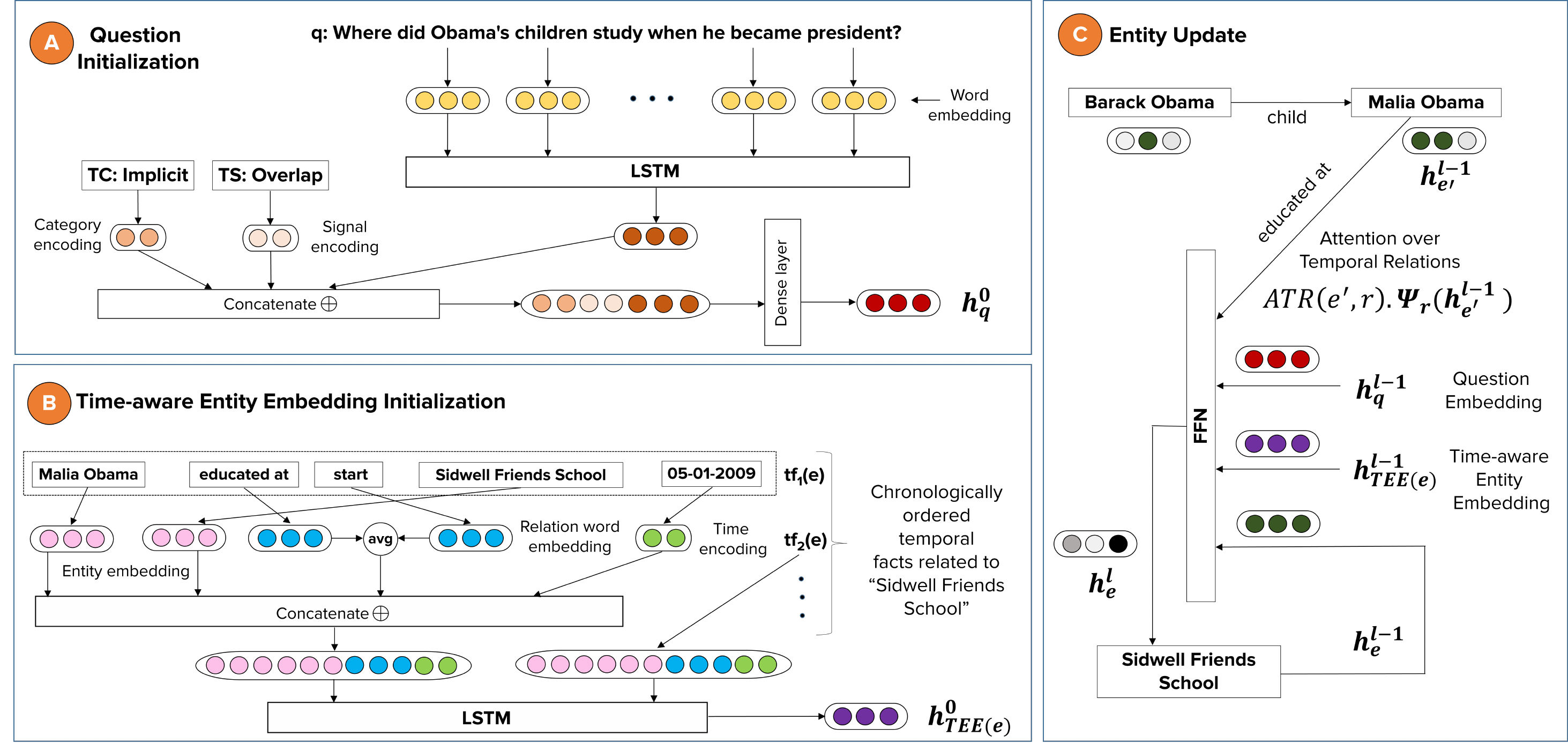}
    \vspace*{-0.3cm}
	\caption{Architecture of the R-GCN model in \exaqt, that includes several signals of temporal information.}
	\label{fig:rgcn}
	\vspace*{-0.3cm}
\end{figure*}


\myparagraph{R-GCN basics} The answer prediction method of \exaqt is inspired by the Relational Graph Convolution Network model~\cite{schlichtkrull2018modeling},
an extension of GCNs~\cite{duvenaud2015convolutional}
tailored for handling large-scale
relational data such as knowledge graphs. Typically, a GCN convolves the features (equivalently, representations or embedding vectors) of nodes belonging to a local neighborhood and propagates them to their nearest neighbors. The learned entity representations are used in 
node classification. Here, this classification decision is whether a node is an answer to the input question or not.
%
%

In this work, we use the widely popular \graftnet model~\cite{sun2018open} that adapted R-GCNs to deal with heterogeneous QA over KGs and text~\cite{bhutani2019online,oguz2021unified}.
In order to apply such a mechanism for answer prediction in our setup, we convert our answer graph from the previous step into a \textit{directed relational graph} and build upon the $KG$-only setting of \graftnet.
In a relational graph, entities, literals, and types become nodes, while predicates (relations) become edge labels.
Specifically, we use the 
KG RDF dump that 
contains normal SPO triples for binary facts by employing
reification~\cite{hernandez2015reifying}. 
Reified triples can then be straightforwardly represented as a directed relational graph~\cite{sun2018open}.
\exaqt introduces four major extensions
over the R-GCN in \graftnet
to deal with the task of temporal QA: 
\squishlist
    \item we embed temporal facts to enrich representations of entity nodes, creating \textit{time-aware entity embeddings} (TEE);
    \item we encode \textit{temporal question categories} (TC) and \textit{temporal signals} (TS) to enrich question representations;
    \item we employ \textit{time encoding} (TE) to obtain the vector representations for timestamps;
    \item we propose \textit{attention over temporal relations} (ATR) to distinguish the same relation but with different timestamps as objects.
\squishend

\noindent In the following, we describe how we encode and update the node representations and perform answer prediction in our 
extended
R-GCN architecture for handling temporal questions. Our neural architecture is shown in Fig.~\ref{fig:rgcn}, while Table~\ref{tab:notation} summarizes notation for the salient concepts used in this phase.

\vspace*{-0.3cm}
\subsection{Question representation}
\label{subsec:quesrep}

\subsubsection{Initialization}
\label{subsubsec:qrepinit}

To encode a temporal question, we first determine its temporal category and extract temporal signals (Sec.~\ref{sec:concepts}). 

\myparagraph{Temporal category encoding (TCE)} We adopt a noisy yet effective strategy for labeling categories for temporal questions, and leave more sophisticated (multi-label) classification as future work. We use a four-bit multi-hot (recall that a question can belong to multiple categories) vector where each bit indicates whether the question falls into that category. Our tagger works as follows:
\squishlist
    \item A question is tagged with the \phrase{EXPLICIT} category if the annotators SUTime~\cite{chang2012sutime} or HeidelTime~\cite{strotgen2010heideltime} detect an explicit temporal expression inside it;
    \item A question is tagged with the \phrase{IMPLICIT} category if it contains any of the temporal signal words (we used the dictionary compiled by~\cite{setzer2002temporal}), and satisfies certain part-of-speech patterns;
    \item A question is of type \phrase{TEMPORAL ANSWER} if it starts with phrases like \phrase{when ...}, \phrase{in which year ...}, and \phrase{on what date ...};
    \item A question is tagged with the \phrase{ORDINAL} category if it contains an ordinal tag as labeled by the Stanford CoreNLP system~\cite{angeli2015leveraging}, along with certain keywords and part-of-speech patterns.
\squishend

\myparagraph{Temporal signal encoding (TSE)} There are $13$ temporal relations defined in Allen's interval algebra for temporal reasoning~\cite{allen1983maintaining}, namely: \phrase{equals}, \phrase{before}, \phrase{meets}, \phrase{overlaps}, \phrase{during}, \phrase{starts}, and \phrase{finishes}, with respective inverses for all of them except \phrase{equals}. We simplify these relations and adapt the strategy in~\cite{jia2018tempquestions} into $7$ broad classes of temporal signals:
\squishlist
\item \phrase{before} and \phrase{meets} relations are treated as \phrase{BEFORE} signals;
\item \phrase{before-inverse} and \phrase{meet-inverse} relations are collapsed into \phrase{AFTER} signals;
\item \phrase{starts} and \phrase{finishes} relations are respectively mapped to \phrase{START} and \phrase{FINISH} signals;
\item words with ordinal tags and \phrase{last} are mapped to \phrase{ORDINAL}; 
\item all other relations are treated as \phrase{OVERLAP} signals;
\item absence of any signal word triggers the \phrase{NO SIGNAL} case.
\squishend
We map signal words to temporal signals in questions using a dictionary.
We then encode these signals using a $7$-bit (a question can contain multiple signals) vector,
where each bit indicates the presence or absence of a particular
temporal signal. 

Along with these temporal categories and temporal signals, we use a Long Short-Term Memory Network (LSTM) 
to model the 
words in the question as a sequence (see block A in Fig.~\ref{fig:rgcn}). Overall, we represent a question $q$ with $|q|$ words as:
\begin{equation}
  \boldsymbol{h^{0}_{q}}=FFN({\boldsymbol{TCE(q)}} \oplus {\boldsymbol{TSE(q)}} \oplus LSTM(\boldsymbol{w_{1}},...,\boldsymbol{w_{|q|}})) 
\end{equation}
Here $\boldsymbol{TCE(q)}$ and $\boldsymbol{TSE(q)}$ 
are
multi-hot vectors encoding the temporal categories and temporal signals 
present in $q$, and $\boldsymbol{w_i}$ represent the pre-trained word embeddings (from Wikipedia2Vec~\cite{yamada2020wikipedia2vec}) of the $i^{th}$ word in $q$. 
We concatenate ($\oplus$) the $\boldsymbol{TCE(q)}$ and $\boldsymbol{TSE(q)}$ vectors with the output vector from the final state of the LSTM. Finally, we pass this concatenated vector through
a Feed Forward Network (FFN) and obtain the initial embedding of $q$, denoted as $\boldsymbol{h^0_q}$.

\subsubsection{Update}

In subsequent layers, the embedding of the question gets updated with the embeddings of the entities belonging to it (i.e. the question entities obtained from NERD) as follows:
\begin{equation}
  \boldsymbol{h^l_q}=FFN(\sum_{e \in NERD(q)} \boldsymbol{h^{l-1}_e})
\end{equation}
where $NERD(q)$ contains the entities for question $q$ and $\boldsymbol{h^{l-1}_e}$ denotes the embedding of an entity $e$ at layer $l-1$. 

\subsection{Entity representation}
\subsubsection{Initialization}

For initializing each entity $e$ in the relational graph, we use fixed-size
pre-trained embeddings $\boldsymbol{x_e}$, also
from
Wikipedia2Vec~\cite{yamada2020wikipedia2vec}.
Along with conventional skip-gram and context models, Wikipedia2Vec utilizes the Wikipedia link graph that learns entity embeddings by predicting neighboring 
entities in the Wikipedia graph, producing more reliable entity embeddings:
\begin{equation}
  \boldsymbol{h^0_e}=\boldsymbol{x_e}
\end{equation}

\subsubsection{Update}

Prior to understanding the update rule for the entities in subsequent layers, we need to introduce the following concepts:
(i) Time encoding (TE); (ii) Time-aware entity embeddings (TEE); 
and (iii) Attention over temporal relations (ATR).

\myparagraph{Time encoding (TE)} Time as an ordering sequence has an inherent similarity to positions of words in text: we thus employ a sinusoidal position encoding method~\cite{vaswani2017attention,zhang2020temporal} to represent a timestamp $ts$.
Here, the $k^{th}$ position (day, month, etc.) in $ts$ will be encoded as:
\begin{equation}
TE(k,j)=
\begin{cases}
      \sin(k / {10000^{\frac{2i}{d}}}), & \text{if}\ j=2i \\
      \cos(k / {10000^{\frac{2i}{d}}}), & \text{if}\ j=2i+1
\end{cases}
\end{equation}
where $d$ is the dimension of the time encoding and $j$ is the (even/odd) position in the $d$-dimensional vector. 
Further, we represent $\boldsymbol{TE(ts)}$, i.e. the time encoding of $ts$, as the summation of the encodings of each of its corresponding positions.
This time encoding method provides an unique encoding to each timestamp and ensures \textit{sequential ordering} among the timestamps~\cite{zhang2020temporal}, that is vital for reasoning signals like \textit{before} and \textit{after} in temporal questions.

\myparagraph{Time-aware entity embedding (TEE)}
An entity $e$ present in the relational graph is associated with a number of temporal facts
$tf^e_1, tf^e_2, ... tf^e_n$
(Sec.~\ref{sec:concepts}) in our answer graph. 
A temporal fact $tf^e$ is said to be associated with an entity $e$ if $e$ is present in any position of the fact (subject, object or qualifier object).
We encode each $tf^e$ as the concatenation of its entity embeddings, relation embeddings (averaged) and time encodings of the timestamps (as shown in block B of Fig.~\ref{fig:rgcn}). Further, we arrange each fact in $\{tf^e\}$ in a chronological order and pass them through
an LSTM network. Finally, the output from the final state of the LSTM can be used as
the
time-aware entity representation of $e$, TEE($e$), that is vital for reasoning through the R-GCN model: 
\begin{equation}
  \boldsymbol{h^{0}_{TEE(e)}}=LSTM(\boldsymbol{h^{0}_{tf^e_1}}, \boldsymbol{h^0_{tf^e_2}}, ...,\boldsymbol{h^0_{tf^e_n}}) 
\end{equation}
In subsequent layers, the embedding of $TEE(e)$ will be updated as the embeddings of its constituent entities get updated.

\myparagraph{Attention over temporal relations (ATR)}
In temporal QA, we need to distinguish entities associated with the same relation but having different timestamps (facts with same temporal predicate but different objects, like several \struct{educated at} facts for a person). We thus introduce the concept of temporal attention here, adapting the more general notion of attention over relations in \graftnet~\cite{sun2018open}.

While computing temporal attention over a relation $r$ connected with entity $e$, we concatenate the corresponding relation embedding with the time encoding 
of its timestamp object and compute its similarity with the question embedding at that stage:
\begin{equation}
  ATR(e, r)=softmax({\boldsymbol{x_{r}}\oplus \boldsymbol{TE({ts}_r)}}^{T} \boldsymbol{h^{(l-1)}_{q}})
\end{equation}  
where the softmax normalization is over all outgoing edges from $e$, $\boldsymbol{x_r}$ is the pre-trained relation vector embedding for 
relation $r$ (Wikipedia2Vec embeddings averaged over each word of the KG predicate), and $\boldsymbol{TE({ts}_r)}$ is the time encoding of the timestamp associated with the relation $r$. 
For relations not connected with any timestamp, we use a random vector
for $\boldsymbol{TE({ts}_r)}$.

\myparagraph{Putting it together} We are now in a position to specify the update rule for entity nodes which involves a single-layer FFN over the concatenation of 
the following four states (see block C of Fig.~\ref{fig:rgcn}): 
\begin{equation}
 \boldsymbol{h^{l}_{e}}=FFN\left(\left[ \begin{array}{ccc}
 \boldsymbol{h^{l-1}_{e}}&\\
 \boldsymbol{h^{l-1}_{q}}& \\
 \boldsymbol{h^{l-1}_{TEE(e)}}& \\
\sum_{r}\sum_{e'\in{nbd_r(e)}}(ATR(e',r).\boldsymbol{\psi_{r}(h^{l-1}_{e'})})& \end{array} \right]\right)
\end{equation}
Here, (i) the first term corresponds to the entity's representation from the previous layer;
(ii) the second term denotes the question's representation from the previous layer;
(iii) the third term denotes the previous layer's representation of the
time-aware entity representation $TEE(e)$; and
(iv) the fourth term aggregates the states from the entity $e$'s neighbors.
In the fourth term, the relation-specific neighborhood $nbd_r$ corresponds to the set of entities connected to $e$ via relation $r$,
$ATR(e',r)$ is the attention over temporal relations, and
$\boldsymbol{\psi_{r}(h^{l-1}_{e'})}$ is the relation-specific transformation depending on the type and direction of an edge: 
\begin{equation}
\boldsymbol{\psi_{r}(h^{l-1}_{e'})}=PPR^{l-1}_{e'} \cdot FFN(\boldsymbol{x_r},\boldsymbol{h^{l-1}_{e'}})
\end{equation}
Here $PPR^{l-1}_{e'}$ is a Personalized PageRank~\cite{haveliwala2003topic} score obtained in the same way as in \graftnet~\cite{sun2018open} to control the propagation of  embeddings along paths starting from the question entities. 

\subsection{Answer prediction}
The final entity representations ($\boldsymbol{h^{l}_{e}}$) obtained at layer $l$, are then used in a binary classification setup to select the answers. For each entity $e$, we define its
probability to be an answer to $q$:
\begin{equation}
    Pr(e \in \{a\}_q | RG_q, q) = \sigma (\boldsymbol{w^T} \boldsymbol{h^{l}_{e}} + \boldsymbol{b})
\end{equation}
where $\{a\}_q$ is the set of ground truth answers for question $q$, ${RG}_q$ is the relational graph built for answering $q$ from its answer graph, and $\sigma$ is the sigmoid activation function. 
$\boldsymbol{w}$ and $\boldsymbol{b}$ are respectively the weight and bias vectors corresponding to the classifier which is trained using binary cross-entropy loss over these $Pr$ probabilities.

\begin{table} [t]
	\centering
	\resizebox*{\columnwidth}{!}{
		\begin{tabular}{l l}
			\toprule
			\textbf{Notation}				& \textbf{Concept}										        \\ \toprule
		    $\boldsymbol{h^{l}_{e}}$                     & Representation of entity $e$ at layer $l$                     \\
		    $\boldsymbol{h^{l}_{q}}$                     & Representation of question $q$ at layer $l$                   \\
            $\boldsymbol{TCE(q)}$                        & Temporal category encoding for question $q$                   \\
			$\boldsymbol{TSE(q)}$                        & Temporal signal encoding for question $q$                     \\
		    $NERD(q)$                       & Question entities obtained from NERD                          \\
		    $\boldsymbol{x_e}$, $\boldsymbol{x_r}$                    & Pre-trained entity ($e$) and relation ($r$) embeddings        \\
		    $\boldsymbol{TE(ts)}$                        & Time encoding for timestamp $ts$                              \\
            $tf^e_1, tf^e_2, \ldots$        & Chronologically ordered temporal facts for $e$                \\		    
			$\boldsymbol{h^{l}_{tf^e_i}}$                & Representation of the $i^{th}$ temporal fact for $e$ at $l$   \\
			$\boldsymbol{h^{l}_{TEE(e)}}$                & Time-aware entity representation of $e$ at $l$                \\
			$ATR(e, r)$                     & Attention over temporal relation $r$ connected with $e$       \\    
            $\boldsymbol{\psi_{r}(h^{l}_{e})}$           & Relation $r$-specific transformation of $h^{l}_{e}$           \\
			$PPR^{l}_{e}$                   &  Personalized PageRank score for entity $e$ at $l$            \\ \bottomrule
	\end{tabular}}
	\caption{Notation for concepts in the R-GCN of \exaqt.}
	\label{tab:notation}
	\vspace*{-0.7cm}
\end{table}
\section{Experimental Setup}
\label{sec:setup}

\subsection{Benchmark}
\label{subsec:benchmark}

\begin{table} [t]
	\centering
\resizebox{\columnwidth}{!}{
    \begin{tabular}{l r r r r r}
       \toprule     
        \textbf{Category}                                   & \textbf{Explicit}	& \textbf{Implicit}	& \textbf{Temp. Ans.}   & \textbf{Ordinal} & \textbf{Total}	\\ \toprule
        \textbf{Free917}~\cite{cai2013large}                & $44$	            & $4$	            & $76$ 	                & $11$              & $135$         \\
        \textbf{WebQ}~\cite{berant2013semantic}     & $315$	            & $77$ 	            & $283$                 & $113$             & $788$         \\
        \textbf{ComplexQ}~\cite{bao2016constraint}	& $217$	            & $131$	            & $43$ 	                & $33$              & $424$         \\
        \textbf{GraphQ}~\cite{su2016generating}	& $264$	            & $30$	            & $13$ 	                & $42$              & $349$         \\
        \textbf{ComplexWebQ}~\cite{talmor2018web}	& $1356$	            & $224$	            & $595$ 	                & $315$              & $2490$         \\
        \textbf{ComQA}~\cite{abujabal2019comqa}             & $669$             & $355$             & $1180$                & $1587$            & $3791$        \\
        \textbf{LC-QuAD}~\cite{trivedi2017lc}	            & $122$             & $19$              & $0$                   & $26$              & $167$         \\
        \textbf{LC-QuAD 2.0}~\cite{dubey2019lc}             & $3534$            & $636$	            & $3726$                & $819$             & $8715$        \\ \midrule
        \textbf{Total}	                                    & $6521$            & $1476$            & $5916$ 	            & $2946$            & $16859$   \\ \bottomrule
    \end{tabular}}
    \caption{Distribution of question types by source in \timequestions. The sum $16859$ exceeds
   the number of questions $16181$ as some questions belong to multiple categories.}
  \label{tab:data}
  \vspace*{-0.7cm}
\end{table}

Previous collections on temporal questions, \textsc{TempQuestions}~\cite{jia2018tempquestions} and \textsc{Event-QA}~\cite{DBLP:conf/cikm/CostaGD20} contain only about a thousand questions each, and are not suitable for building neural models. We leverage recent community efforts in QA benchmarking,
and we search through eight KG-QA datasets for time-related questions. The result is a new compilation, \timequestions, with $16,181$ questions, that we release with this paper (details in Table~\ref{tab:data}). 
Since some of these previous benchmarks were over Freebase or DBpedia, we used Wikipedia links in these KGs to map them to Wikidata, the largest and most actively growing public KG today, and the one that we use in this work.
Questions in each benchmark are tagged for temporal expressions using SUTime~\cite{chang2012sutime} and HeidelTime~\cite{strotgen2010heideltime}, and for signal words using a dictionary compiled by~\cite{setzer2002temporal}. Whenever a question is found to have at least one temporal expression or signal word, it becomes a candidate temporal question. This candidate set (ca. $20k$ questions) was filtered for false positives by the authors.
For each of these questions, the authors manually verified the correctness of the answer, and if incorrect, replaced it with the right one.
Moreover, each question is manually tagged with its temporal question category (explicit, implicit, temporal answer, or ordinal) that may help in building automated classifiers for temporal questions, a sub-problem interesting in its own right. We split our benchmark in a $60:20:20$ ratio for creating the training ($9708$ questions), development ($3236$) and test ($3237$) sets.
%
%

\subsection{Baselines}
\label{subsec:baselines}

We use the following recent methods for complex KG-QA as baselines to compare \exaqt with. All baselines were trained and fine-tuned using the train and dev sets of \timequestions, respectively. They are the most natural choice of baselines as \exaqt is inspired by components in these methods for building its pipeline: while \uniqorn~\cite{pramanik2021uniqorn} showed the effectiveness of GSTs in complex KG-QA, \graftnet~\cite{sun2018open} and \pullnet~\cite{sun2019pullnet} showed the value of R-GCNs for answer prediction. These techniques are designed for dealing with heterogeneous answering sources (KGs and text), and we use their KG-only variants:
\squishlist
\item \uniqorn~\cite{pramanik2021uniqorn}: This is a method for answering complex questions using Group Steiner Trees, and is an extension of~\cite{lu2019answering};
\item \graftnet~\cite{sun2018open}: This was the first technique to adapt R-GCNs for QA over heterogeneous sources; 
\item \pullnet~\cite{sun2019pullnet}: This algorithm extended the \graftnet classifier to the scenario of multi-hop questions.
We used a reimplementation as the code is not public.
\squishend

\subsection{Metrics}
\label{subsec:metrics}

All systems return a ranked list of answers, consisting of KG entities or literals associated with unique identifiers.
We thus use the following metrics for evaluating
\exaqt and the baselines,
averaged over questions in the benchmark:
\squishlist
    \item P@1: Precision at the top rank is one if the highest ranked answer is correct, and zero otherwise.
    \item MRR: This is the reciprocal of the first rank where we have a correct answer. If the correct answer does not feature in the ranked list, MRR is zero.
    \item Hit@5: This is set to one if a correct answer appears in the first five positions, and zero otherwise. 
\squishend

\subsection{Initialization}
\label{subsec:init}

\myparagraph{Configuration} We use the Wikidata KG dump (\url{https://dumps.wikimedia.org/wikidatawiki/entities/}) in NTriples format from April 2020, comprising $12B$ triples and taking $2$ TB when uncompressed on disk. We subsequently removed language tags, external IDs, schema labels and URLs from the dump, leaving us with about $2B$ triples with $340$ GB disk space consumption.

For BERT fine-tuning, positive and negative instances were created from the \timequestions train and dev sets in the ratio $1:5$. These instances were combined and split in the ratio $80:20$ (test set not needed), where the first split was used for training and the second for hyperparameter selection, respectively, for BERT fine-tuning. We use the BERT-base-cased model for sequence pair classification (\url{https://bit.ly/3fRVqAG}).
Best parameters for fine-tuning were: accumulation $= 512$, number of epochs $= 2$, dropout $= 0.3$, mini-batch size $= 50$ and weight decay $= 0.001$. We use AdamW
as the optimizer with a learning rate of $3 \times 10^{-5}$. During answer graph construction, we use top-$25$ 
question-relevant facts ($|\{f_{qrel}\}| = 25$), top-$25$ GSTs ($k = 25$), and top-$25$ temporal facts ($|\{tf_{qrel}\}| = 25$).

\myparagraph{R-GCN model training}
100-dimensional embeddings for question words, relation (KG predicate) words and entities, are obtained from Wikipedia2Vec~\cite{yamada2020wikipedia2vec}, and learned from the Wikipedia dump of March 2021. Dimensions of TCE, TSE, TE and TEE (Sec.~\ref{sec:method-2}) were all set to $100$ as well. The last hidden states of LSTMs were used as encodings wherever applicable. This was trained on an Nvidia Quadro RTX 8000 GPU server. Hyperparameter values were tuned on the \timequestions dev set: number of GCN layers  = $3$, number of epochs  = $100$, mini-batch size = $25$, gradient clip = $1$, learning rate = $0.001$, LSTM dropout = $0.3$, linear dropout  = $0.2$, and fact dropout  = $0.1$. The ReLU activation function was used. 

\vspace*{-0.3cm}
\section{Key findings}
\label{sec:results}

\begin{table*} [t] \small
	\newcolumntype{G}{>{\columncolor [gray] {0.90}}c}
	\resizebox{\textwidth}{!}{
	\begin{tabular}{l G G G c c c G G G c c c G G G}
		\toprule
		\textbf{Category}									&	\multicolumn{3}{G}{\textbf{Overall}}												&	\multicolumn{3}{c}{\textbf{Explicit}}											& 	\multicolumn{3}{G}{\textbf{Implicit}}												& 	\multicolumn{3}{c}{\textbf{Temp. Ans.}}												& 	\multicolumn{3}{G}{\textbf{Ordinal}}												\\ \midrule
		\textbf{Method}									&	\textbf{P@1}			&	\textbf{MRR}			&	\textbf{Hit@5}			&	\textbf{P@1}			&	\textbf{MRR}			&	\textbf{Hit@5}			&	\textbf{P@1}			&	\textbf{MRR}			&	\textbf{Hit@5}			&	\textbf{P@1}			&	\textbf{MRR}			&	\textbf{Hit@5}			&	\textbf{P@1}			&	\textbf{MRR}			&	\textbf{Hit@5}			\\ \toprule
		\textbf{\uniqorn}~\cite{pramanik2021uniqorn}	&	$0.331$					&	$0.409$					&	$0.538$					&	$0.318$					&	$0.406$					&	$0.536$					&	$0.316$					&	$0.415$					&	$0.545$					&	$0.392$					&	$0.472$					&	$0.597$					&	$0.202$					&	$0.236$					&	$0.356$					\\
		\textbf{\graftnet}~\cite{sun2018open}						&	$0.452$					&	$0.485$					&	$0.554$					&	$0.445$					&	$0.478$					&	$0.531$					&	$0.428$	&	$0.465$					&	$0.525$					&	$0.515$	&	$0.568$					&	$0.660$					&	$0.322$					&	$0.313$					&	$0.371$					\\
		\textbf{\pullnet}~\cite{sun2019pullnet}									&	$0.105$	&	$0.136$	&	$0.186$	&	$0.022$	&	$0.043$	&	$0.075$	&	$0.081$					&	$0.123$	&	$0.192$	&	$0.234$					&	$0.277$	&	$0.349$	&	$0.029$	&	$0.049$	&	$0.083$	\\ \midrule
		\textbf{\exaqt}						&	$\boldsymbol{0.565}$*					&	$\boldsymbol{0.599}$*					&	
		$\boldsymbol{0.664}$*				            	&	
		$\boldsymbol{0.568}$*					&	$\boldsymbol{0.594}$*					&	$\boldsymbol{0.636}$*					&	$\boldsymbol{0.508}$*					&	$\boldsymbol{0.567}$*					&	
		$\boldsymbol{0.633}$*					                &	
		$\boldsymbol{0.623}$*					&	
		$\boldsymbol{0.672}$*					&	
		$\boldsymbol{0.756}$*					&	
		$\boldsymbol{0.420}$*					&	$\boldsymbol{0.432}$*					&	$\boldsymbol{0.508}$*					\\ \bottomrule
	\end{tabular}}
	\\ \raggedright Statistical significance of \exaqt over the strongest baseline (\graftnet), under the $2$-tailed paired $t$-test, is marked with an asterisk (*) ($p < 0.05$).
	\caption{Performance comparison of \exaqt with three complex QA baselines over the \timequestions test set.}
	\label{tab:main-res}
	\vspace*{-0.7cm}
\end{table*}

Answering performance of \exaqt and baselines are in Table~\ref{tab:main-res} (best value in column in \textbf{bold}). Main observations are as follows.


\myparagraph{\exaqt outperforms baselines} The main observation from Table~\ref{tab:main-res} is the across-the-board superiority of \exaqt over the baselines. 
Statistically significant results for each category, baseline and metric, indicate that general-purpose complex QA systems are not able to deal with the
challenging
requirements of temporal QA, and that 
temporally augmented methods are needed. Outperforming each baseline offers individual insights, as discussed below.

\myparagraph{GSTs are not enough} GSTs are a powerful mechanism for complex QA that identify backbone skeletons in KG subsets and prune irrelevant information from noisy graphs. While this motivated the use of GSTs as a building block in \exaqt, outperforming the \uniqorn~\cite{pramanik2021uniqorn} method shows that non-terminals (internal nodes) in GSTs, by themselves, are not enough to answer temporal questions.

\myparagraph{Augmenting R-GCNs with time information works well} The fact that R-GCNs are a powerful model is clear from the fact that \graftnet, without any explicit support for temporal QA, emerges as the strongest baseline in this challenging setup. A core contribution of our work is to 
extend
R-GCNs with different kinds of temporal evidence. Improving over \graftnet shows that our multi-pronged mechanism (with TEE, ATR, TCE, TSE, and TE) 
succeeds
in advancing
the scope of R-GCN models to questions with temporal intent. Ablation studies (Sec.~\ref{sec:analysis}) show that each of these ``prongs'' play 
active
roles in the overall performance of \exaqt.

\myparagraph{Not every question is multi-hop} \pullnet is a state-of-the-art system for answering multi-hop chain-join questions (\utterance{where was Obama's father born?}). It may appear strange that \pullnet, offered as an improvement over \graftnet, falls short in our setup. 
Inspecting 
examples makes the reason for this clear: \pullnet has an assumption that all answers are located on a $2$-hop circumference of the question entities
(ideally, $T$-hop, where $T$ is a variable that needs to be fixed for a benchmark: $1$ is an oversimplification, while $3$ is intractable
for a large KG, and hence our choice of $2$ for \timequestions).
When this is not the case (for instance, the slightly tricky situation when an answer is in a qualifier of
a 2-hop fact:
\utterance{when did obama's children start studying at sidwell friends school?} or the question is simple: \utterance{when was obama born?}), \pullnet cannot make use of this training point as it relies on shortest KG paths between question and answer entities. This uniform $T$-hop assumption is not always practical, and does not generalize to 
situations
beyond what \pullnet was trained and evaluated on. 

\myparagraph{Temporal categories vary by difficulty} We use manual ground-truth labels of question categories from our benchmark to drill down on class-wise results (the noisy tagger from Sec.~\ref{subsubsec:qrepinit} has $\simeq 90\%$ accuracy). 
Questions with temporal answers are clearly the easiest. Note that this includes questions starting with \phrase{when}, that many models tackle with dedicated lexical answer types~\cite{bast2015more,abujabal2017automated}, analogous to location-type answers for \phrase{where ...?} questions. Questions with explicit temporal expressions are the next rung of the ladder: while they do require reasoning, explicit years often make this matching easier (\utterance{who became president of south africa in 1989?}). Questions with implicit expressions are more challenging: we believe that this is where the power of R-GCNs truly shine, as GST-based \uniqorn clearly falls short. Finally, questions with temporal ordinals seem to be beyond what implicit reasoning in graph neural networks can handle: with P@1 $< 0.5$, they pose the biggest research challenge. We believe that this calls for revisiting symbolic reasoning, ideally plugged into neural GCN architectures.

\vspace*{-0.1cm}
\section{In-depth analysis}
\label{sec:analysis}

\begin{table} [t] \small
	\centering
	\begin{tabular}{l c c c}
		\toprule
		\textbf{NERD}   &\textbf{Recall} &\textbf{\#Question entities}	\\  \toprule
		\tagme	        & $0.682$               & $2.9$			            \\ 
		\elq		    & $0.716$ 	            & $1.7$ 	                \\
		\aida	        & $0.541$               & $2.8$			            \\ \midrule
		\tagme + \elq   & $\boldsymbol{0.758}$  & $3.5$		                \\
		\aida + \elq	& $0.729$               & $3.5$		                \\
		\tagme + \aida  & $0.701$               & $4.3$                     \\   \bottomrule
	\end{tabular}
	\caption{Comparing various NERD methods on the test set.}
	\label{tab:nerd}
	\vspace*{-0.7cm}
\end{table}

\myparagraph{NERD variants}
We experimented with \tagme~\cite{ferragina2010tagme}, \aida~\cite{hoffart2011robust}, and
\elq~\cite{li2020efficient}, going by the most popular to the most recent choices. Effects
of various choices are in Table~\ref{tab:nerd}. Our best configuration is \tagme + \elq. \tagme (used without
threshold on pruning entities) and \elq (run with default parameters)
nicely complement each other, since one is recall-oriented
(\tagme) and the other precision-biased (\elq). Answer recall measures
the fraction of questions for which at least one gold answer was present in the final
answer graph (test set). \aida + \elq detects a similar number
of entities per question, but is slightly worse w.r.t. answer recall.


\begin{table} [t] \small
	\centering
	\resizebox*{\columnwidth}{!}{
	\begin{tabular}{l c c}
		\toprule
		\textbf{Step in \exaqt pipeline}                & \textbf{Recall}   & \textbf{\#Candidates}   \\  \toprule
		All KG facts of NERD entities		            & $0.758$ 	        & $2491$       	        \\
		Facts selected by BERT 	                        & $0.719$          	& $48$                  \\ 
		Shortest paths injected for connectivity	    & $0.720$        	& $49$                  \\ 
		GSTs on largest component                       & $0.613$           & $13$                  \\
		Union of GSTs from all components 	            & $0.640$           & $14$	                \\
		Completed GSTs from all components              & $0.671$           & $21$                  \\   
		Temporal facts added by BERT                    & $0.724$           & $67$                  \\   \bottomrule
	\end{tabular}}
	\caption{Understanding the recall-oriented Stage 1 of \exaqt.}
	\label{tab:error}
	\vspace*{-0.7cm}
\end{table}

\myparagraph{Understanding Stage 1} Traversing over the steps in the recall-oriented graph construction phase of \exaqt, we try to understand where we gain (and lose) answers to temporal questions (Table~\ref{tab:error}, test set). First, we see that even two NERD systems cannot guarantee perfect answer recall ($75.8\%$). The fall from Row 1 to 2 is expected, as one cannot compute graph algorithms efficiently over such large graphs as induced by all facts from Row 1. Adding shortest paths (Row 3), while making the answer graph more connected (before: $1.58$ connected components per question, after: $1.16$), also marginally helps in bringing correct answers into the graph. From Rows 4 and 5, we see that taking a union of top-$k$ ($k = 25$) GSTs from \textit{each} connected component proves worthwhile (increase from 0.613 to 0.640), and so does \textit{completing} the GSTs (further rise to $0.671$). Finally, adding temporal facts provides a critical boost, taking the answer recall at the end of Stage 1 to a respectable $72.4\%$. This translates to $2343$ questions having answers in the graph passed on to the R-GCN (cf. $1989$ answers are present in the PPR-based answer graph of \graftnet), out of which $1830$ are answered correctly at the end. The second column, that counts the average number of entities and literals in the answer graph (answer candidates) is highly insightful to get an idea of the graph size at each step, and its potential trade-off with respect to answer recall.

\begin{table} [t] 
	\centering
	\newcolumntype{G}{>{\columncolor [gray] {0.90}}c}
	\resizebox{\columnwidth}{!}{
	\begin{tabular}{l c c c c c}
		\toprule
		\textbf{Category}									&	\textbf{Overall}												&	\textbf{Explicit}											& 	\textbf{Implicit}												& 	\textbf{Temp. Ans.}											& \textbf{Ordinal}												\\ \midrule
				\textbf{\exaqt (Full)}						&	
				$\boldsymbol{0.565}$					&	
		$\boldsymbol{0.568}$					&	$\boldsymbol{0.508}$				 &	
		$\boldsymbol{0.623}$									&	
		$\boldsymbol{0.420}$					\\ \midrule
		\textbf{\textsc{Exaqt}  - TCE}		&	
		$0.545$					&	
		$0.556$										&	
		$0.481$				&	
		$0.590$					&
		$0.406$						
		\\
		\textbf{\textsc{Exaqt}  - TSE}						&	$0.543$	&	
		
		$0.545$					&	
	
		$0.465$					&	
			
		$0.598$					&	
		
		$0.411$		
		\\
		\textbf{\textsc{Exaqt}  - TEE}						&
		
		$0.556$	&	
		
		$0.564$					&		
		$0.475$					&		$0.614$					&	$0.413$	
		\\ 
		\textbf{\textsc{Exaqt}  - TE}									&	$0.553$						&	$0.556$					&	$0.495$						&	$0.613$								&	$0.398$						\\
		\textbf{\textsc{Exaqt}  - ATR}						&	$0.534$					&	
		$0.527$					&		$0.465$					&		$0.594$					&	
		$0.411$				\\ \bottomrule
	\end{tabular}}
	\caption{Inspecting the precision-oriented Stage 2 of \exaqt.}
	\label{tab:ablation}
	\vspace*{-0.7cm}
\end{table}

\myparagraph{Understanding Stage 2} We 
performed \textit{ablation} studies to understand the relative influence of the individual temporal components in the precision-oriented Stage 2 of \exaqt: the R-GCN answer classifier.
Table~\ref{tab:ablation} shows P@1 results on the test set, where the full model achieves the best results overall and also for each category.
The amount of drop from the full model (Row 1) indicates the degree of importance of a particular component. The most vital enhancement
is the attention over temporal relations (ATR).
All other factors offer
varying degrees of assistance.
An interesting observation is that TCE, while playing a moderate role in most categories, is of the highest importance for questions with temporal answers: even knowing that a question belongs to this category helps the model.

\begin{table} [t] \small
	\centering
		\begin{tabular}{l}
			\toprule
            \utterance{what did abraham lincoln do before he was president?}  \\
            \utterance{who was the king of troy when the trojan war was going on?}       \\
             \utterance{what films are nominated for the oscar for best picture in 2009?}\\
            \utterance{where did harriet tubman live after the civil war?}\\
             \utterance{when did owner bill neukom's sports team last win the world series?}\\
            \bottomrule
	\end{tabular} 
	\caption{Anecdotal examples that \exaqt answered correctly.} 
	\label{tab:anecdotes}
	\vspace*{-0.9cm}
\end{table}

\myparagraph{Anecdotal examples}
Table~\ref{tab:anecdotes} shows samples
of test questions that are 
successfully processed by
\exaqt but none of the baselines.

\section{Related Work}
\label{sec:related}


\myparagraph{Temporal QA in IR} 
Supporting temporal intent in query and document processing has been a long-standing research topic in IR~\cite{setzer2002temporal,campos2014survey,kanhabua2016temporal,alonso2011temporal,berberich2010language,navarro2015combining}. This includes work inside the specific use case of QA over text~\cite{harabagiu2005question,saquete2009enhancing,ahn2006supporting,lloret2011text}. Most of these efforts require significant preprocessing and markup of documents.
There is also onus on questions to be formulated in specific ways so as to conform to carefully crafted parsers. These directions often fall short of realistic settings on the Web, where documents and questions are both formulated ad hoc. Moreover, such corpus markup unfortunately does not play a role in structured knowledge graphs. Notable effort in temporal QA includes work of 
\cite{saquete2009enhancing}, which decompose complex questions into simpler components, and recompose answer fragments into responses that satisfy the original intent. Such approaches have bottlenecks from parsing issues.
\exaqt
makes no assumptions on how 
questions are formulated.

\myparagraph{Temporal QA over KGs}
Questions
with temporal conditions
have not received much attention in the KG-QA literature.
The few works that specifically address temporal questions include \cite{jia18tequila,DBLP:conf/cikm/CostaGD20,DBLP:journals/fi/WuZLZG20}. Among these, \cite{jia18tequila} relies on hand-crafted rules with limited generalization, whereas \exaqt 
is automatically trained with distant supervision and covers a much wider territory of questions.
\cite{DBLP:conf/cikm/CostaGD20} introduces the task of event-centric QA, which overlaps with our notion of temporal questions, and introduces a benchmark collection. 
\cite{DBLP:journals/fi/WuZLZG20} presents
a key-value memory network to include KG information about time into a QA pipeline. The method is geared for simple questions, as present in the WebQuestions benchmark.



\myparagraph{Temporal KGs} Of late, understanding large KGs as a dynamic body of knowledge has gained attention, giving rise to the notion of temporal knowledge graphs or temporal knowledge bases~\cite{dhingra2021time,trivedi2017know}. Here, each edge (corresponding to a fact) is associated with a temporal scope or validity~\cite{leblay2018deriving}, with current efforts mostly focusing on the topic of temporal KG completion~\cite{garcia2018learning,lacroix2020tensor,goel2020diachronic}. A very recent approach has explored QA over such temporal KGs, along with
the creation of an associated benchmark~\cite{saxena2021question}. 

\section{Conclusions}
\label{sec:conclusion}

%
%
Temporal questions have been underexplored in QA, and so has temporal information in KGs,
despite their importance for knowledge workers like analysts or journalists as well as advanced information needs of lay users. 
This work on the \exaqt method has presented a complete pipeline for filling this gap, based on a judicious combination of BERT-based classifiers and graph convolutional networks. Most crucially, we devised new methods for augmenting these components with temporal signals. Experimental results with a large collection of complex temporal questions demonstrate the superiority of \exaqt over state-of-the-art
general-purpose
methods for
QA over knowledge graphs.

\vspace*{0.2cm}
\myparagraph{Acknowledgements} We thank Philipp Christmann and Jesujoba Alabi from the MPI for Informatics for useful inputs at various stages of this work. Zhen Jia was supported by (i) China Academy of Railway Sciences Corporation Limited (2019YJ106); and (ii) Sichuan Science and Technology Program (2020YFG0035). 
\clearpage

\bibliographystyle{ACM-Reference-Format}
\balance
\bibliography{exaqt}

\end{document}